\documentclass[%
floatfix,
 aip,
 amsmath,amssymb,
 reprint,%
]{revtex4-1}

\usepackage{graphicx}% Include figure files
\usepackage{subfig}
\usepackage{dcolumn}% Align table columns on decimal point
\usepackage{bm}% bold math
\usepackage[utf8]{inputenc}
\usepackage[T1]{fontenc}
\usepackage{mathptmx}
\usepackage{etoolbox}

\usepackage{chemformula} % Formula subscripts using \ch{}
\usepackage[T1]{fontenc} % Use modern font encodings
\usepackage[version=3]{mhchem}
\usepackage{enumerate}
\usepackage{amsmath}
\usepackage{mathrsfs}
\usepackage{amssymb}
\usepackage{subfig}
\usepackage{amssymb}
\usepackage{stackrel}
\usepackage{amscd}
\usepackage{physics}
\usepackage{mciteplus}
\usepackage{natbib}
\usepackage{array}
\usepackage{booktabs}
\usepackage{listings}
\usepackage{lmodern}
\usepackage{mathpazo}
\usepackage{microtype}
\usepackage{float}
\usepackage{caption}
\usepackage{setspace}
\usepackage{xkeyval}
\usepackage{color}
\usepackage{siunitx}
\usepackage{atbegshi} %% solves the \box255 problem
\usepackage{amsmath,amssymb}
\newcommand{\cket}[1]{\left| #1 \right)} % for non-Hermitian Dirac kets
\newcommand{\cmatrixel}[3]{\left( #1 \vphantom{#2#3} \right|
 #2 \left| #3 \vphantom{#1#2} \right)} % for Dirac matrix elements
\newcommand{\cbraket}[2]{\left( #1 \vphantom{#2} \vert 
 #2 \vphantom{#1} \right)} % for quantum Dirac scalar prod.

\makeatletter
\def\@email#1#2{%
 \endgroup
 \patchcmd{\titleblock@produce}
  {\frontmatter@RRAPformat}
  {\frontmatter@RRAPformat{\produce@RRAP{*#1\href{mailto:#2}{#2}}}\frontmatter@RRAPformat}
  {}{}
}%
\makeatother

\draft

\begin{document}

%\preprint{AIP/123-QED}

\title[CAP Analysis and Optimization]{Analysis and Optimization of Resonance Energies and Widths Using Complex Absorbing Potentials}
% Force line breaks with \\
\author{J.A Gyamfi}
 \email{jerrymanappiahene.gyamfi@kuleuven.be}
 \affiliation{Department of Chemistry, KU Leuven, Celestijnenlaan 200F, B-3001 Leuven, Belgium}%Lines break automatically or can be forced with \\
\author{T.-C. Jagau}%
 \email{thomas.jagau@kuleuven.be}
\affiliation{Department of Chemistry, KU Leuven, Celestijnenlaan 200F, B-3001 Leuven, Belgium}%

\date{\today}% It is always \today, today,
             %  but any date may be explicitly specified

\begin{abstract}
Complex absorbing potentials (CAPs) are artificial potentials added to electronic Hamiltonians to make the wavefunction of metastable electronic states square-integrable. This makes the electronic structure problem of electronic resonances comparable to that of electronic bound states, thus reducing the complexity of the problem. CAPs depend on two types of parameters: the coupling parameter $\eta$ and a set of spatial parameters which define the onset of the CAP. It has been a common practice over the years to minimize the CAP perturbation on the physical electronic Hamiltonian by running an $\eta-$trajectory, whereby one fixes the spatial parameters and varies $\eta$. The optimal $\eta$ is chosen according to the minimum log-velocity criterion. But the effectiveness of an $\eta-$trajectory strongly depends on the values of the fixed spatial parameters. \\
In this work, we propose a more general criterion, called the $\xi-$criterion, which allows one to minimize any CAP parameter, including the CAP spatial parameters. Indeed, we show that fixing $\eta$ and varying the spatial parameters according to a scheme (i.e., running a spatial trajectory) is a more efficient and reliable way of minimizing the CAP perturbations (which is assessed using the $\xi-$criterion). We illustrate the method by determining the resonance energy and width of the temporary anion of dinitrogen, at the Hartree-Fock and EOM-EA-CCSD levels, using two different types of CAPs: the box- and the smooth Voronoi-CAPs.
\end{abstract}

% insert suggested PACS numbers in braces on next line

\maketitle %\maketitle must follow title, authors, abstract and \pacs

% Body of paper goes here. Use proper sectioning commands. 
% References should be done using the \cite, \ref, and \label commands

\section{Introduction}
Owing to their role in diverse fields of industry and science (including biochemistry\cite{boudaiffa00,simons07}, astrophysics\cite{millar17}, semiconductor manufacturing\cite{stoffels01}), research on temporary anions\cite{herbert15,ingolfsson19} (i.e., metastable electronic resonances, with a lifetime in the range of femto- to milliseconds, formed when a neutral molecule captures an electron) has seen a significant increase in the last decades\cite{moiseyev11,jagau22,jagau17}.
It is, however, extremely challenging to study temporary anions (TAs) by means of conventional quantum chemistry methods designed for bound electronic states because the electronic states of TAs are embedded in the continuum of the electronic Hamiltonian. 

Complex-variable techniques offer a way to study TAs. Here, the objective is to make diverging electronic resonance wavefunctions square-integrable\cite{moiseyev11}, making it then possible to apply computational techniques similar to those previously developed for electronic bound states. This class of methods offers techniques like complex-scaling\cite{balslev1971, simon1972, moiseyevRev98, moiseyev11} (where electronic coordinates in the Hamiltonian are scaled by a complex number), complex basis functions \cite{mcCurdyRescigno1978, moiseyevCorcoran79, WhiteMHG15} (where one employs Gaussian basis functions with complex-scaled exponents) and complex absorbing potentials\cite{jolicard86, riss93, zuev14} (where a complex potential is added to the original Hamiltonian). This suite of methods also come with the important advantage that explicit representation of the continuum part of the Hamiltonian spectrum is no longer necessary. Nonetheless, the transformations associated with these techniques render the Hamiltonian non-Hermitian. Consequently, one has to work in the framework of non-Hermitian quantum mechanics \cite{moiseyev11}, where, i) the usual Hilbert-Schmidt inner product is replaced with the so-called \textit{c-product}\cite{moiseyev11, moiseyevCertain78}, and ii) the expectation value of operators like the Hamiltonian become, in general, complex-valued. % Despite these inconveniences, complex-variable techniques are relatively easy to combine with existing bound-state methods. 
In the following, we shall use  $\cket{\bullet}$ to signal we are using the c-product, namely
			\begin{equation}
			\cbraket{\psi_m}{\psi_m} = \int \psi_n \psi_m \ d\vartheta \ ,
			\end{equation}
where $\vartheta$ is the integration variable. 

\par Among the complex-variable techniques, the complex absorbing potential (CAP) method is one of the most widely employed to study TAs. Here, the electronic Hamiltonian $H$ is augmented by a non-Hermitian operator:
		\begin{equation}
		\label{eq:CAP_Hamiltonian}
		H_{CAP} = H -i \eta W \ ,
		\end{equation}
where $\eta$ is a real scalar, and $W$ is a Hermitian operator. The associated eigenvalue problem,
		\begin{equation}
		H_{CAP} \psi = E \psi
		\end{equation}
is characterized by complex eigenvalues $E$ since $H_{CAP}$ is now non-Hermitian. $E$ may be written as a Siegert energy:
		\begin{equation}
		E = E_r - i \Gamma/2
		\end{equation}
where $E_r$ is the energy of the resonance system and $\Gamma$ is related to its lifetime, $\tau$, through the relation $\Gamma = \tau^{-1}$ (here and henceforth, $\hbar = 1$).

Different functional forms of $W$ have been proposed, leading to specific names for the non-Hermitian contribution $-i \eta W$ (referred to as the complex absorbing potential), e.g., the box-CAP\cite{jagau14}, the smooth Voronoi-CAP\cite{sommerfeld15}  and reflection-free CAPs\cite{riss95,riss98,moiseyev98}.

In this work, we shall make use of two types of CAPs: i) quadratic box-CAPs, and ii) smooth Voronoi-CAPs. In both types of CAP, $W$ is a one-electron operator parameterized by a set of spatial parameters which define the spatial boundary dividing space into two regions: one where the CAP is applied and the other where it is not. With a box-CAP, as the name suggests, the spatial boundary in question is a three-dimensional box -- hence, defined by three parameters: $r^o_x, r^o_y, r^o_z$, which are related to the volume of the box. With a quadratic box-CAP, the CAP is quadratic in the electronic coordinates. In the following, we shall refer to quadratic box-CAPs simply as 'box-CAPs'.

With the smooth Voronoi-CAP, the boundary region is constructed from each nucleus' Voronoi cell, with the sharp edges between the cells smoothed out.\cite{sommerfeld15} Here, the operator $W$ depends on a single spatial parameter $r^o$.\cite{sommerfeld15} 

Box-CAPs and smooth Voronoi-CAPs are the most widely used types of CAPs to study TAs. In the Supplementary Material, we give a short overview of the equations governing these two types of CAPs.

To simplify notations, we shall, in general, use the vector $\vb{r}^o$ to indicate spatial parameters (where, for box-CAP, $\vb{r}^o \equiv (r^o_x,r^o_y,r^o_z)$; while for smooth Voronoi-CAP, $\vb{r}^o \equiv r^o$). 

Unlike wavefunctions describing bound electronic states, whereby the wavefunctions vanish in the asymptotic region, the wavefunctions of temporary anions do not. The purpose of a CAP is to make the diverging electronic resonance wavefunction square-integrable by creating an absorbing region where the tail of the wavefunction is artificially damped, enabling thus a practical description of the system using a finite basis set or finite grid points\cite{jolicard86,riss98}. However, the introduction of the unphysical CAP, $-i\eta W$, may lead to a strong perturbation of the original Hamiltonian $H$ in Eq. \eqref{eq:CAP_Hamiltonian} -- therefore, a wrong description of the system as well. For example, in our recent publication\cite{gyamfiJagau22}, where we introduced a computational method (i.e. CAP-AIMD) to simulate the dynamics of temporary anions, we showed how using a wrong CAP may lead to an incorrect description of a temporary anion's molecular dynamics. To obviate such problems, it is imperative to make sure that the CAP is just a small perturbation to the physical electronic Hamiltonian $H$ in Eq. \eqref{eq:CAP_Hamiltonian}. How to achieve this has been an important subject since the advent of the CAP method\cite{kosloff86, neuhauser89A, neuhauser89B, riss93, macias94, riss96, jagau14}.

In practical calculations, where one deals with multidimensional quantum systems, the prevailing way of minimizing CAP reflections is through an optimization of \textit{only} the parameter $\eta$, leaving out the spatial parameters $\vb{r}^o$. There is no consensus on how to choose values for the latter. Not surprisingly, groups of researchers tend to adopt their own convention, mostly based on some empirical evidence. One convention, for example, is to take the spatial parameters $r^o_\alpha$ as $r^o_\alpha = \sqrt{\left<r^2\right>_\alpha}$, where $\left<r^2\right>_\alpha$ is the expectation value of the square of the neutral's electronic coordinate along axis $\alpha \in {x,y,z}$.\cite{jagau14,zuev14,benda17} An alternative scheme is to set $r^o_\alpha= \max_I \abs{R_{I,\alpha}} + c$, where $R_{I,\alpha}$ is the coordinate of the $I-$th nucleus along axis $\alpha$, and $c$ is an arbitrary additive constant.\cite{gayvertBravaya22}. Similar schemes may also be used in setting up the value of $r^o$ for the smooth Voronoi-CAP. 

Once the CAP spatial parameters are set, the criterion usually adopted in choosing $\eta_{opt}$ is the so-called \textit{minimum log-velocity criterion}, and reads as\cite{riss93},
			\begin{equation}
			\label{eq:classic_criterion}
			\eta_{opt} = \min_{\eta} \norm{\eta \frac{dE(\eta')}{d\eta'}\Big\vert_{\eta'=\eta}} \ .
			\end{equation}

The use of $\eta-$trajectories in conjunction with the criterion in Eq. \eqref{eq:classic_criterion} has proved useful in the study of electronic resonances. However, it is undeniable to practitioners of the CAP method that in some cases, the eigenstate related to the  $\eta_{opt}$ chosen through the application of Eq. \eqref{eq:classic_criterion} may be a resonant state with a strong perturbation from the CAP or may correspond to a pseudocontinuum state. In some cases, one fails to locate a resonant state in an $\eta-$trajectory. All these issues stem from the fact that for an $\eta-$trajectory to be useful in locating resonant states and finding optimal values of $\eta$ which produce resonant states with mild perturbations from the CAP, the spatial parameters ought to be proper. In addition, $\eta-$trajectories often present discontinuities, making the application of Eq. \eqref{eq:classic_criterion} not appropriate.

In this paper, our aim is twofold: i) Propose a shift from the minimum log-velocity criterion to a general one called the $\xi-$criterion, which has the expectation value of the CAP, $\left< -i \eta W\right>$, at its center. In this way, it becomes possible to optimize \textit{any} parameter of the CAP, not just $\eta$. Moreover, the possible discontinuities in $\eta-$trajectories  (or the trajectories of any other CAP parameter) cease to be a problem. ii) Show how effective $\vb{r}^o-$trajectories are in minimizing CAP reflections and in finding resonant states. To illustrate this point, we study the $\vb{r}^o-$trajectories for the temporary anion of dinitrogen at the X-CAP-HF/cc-PVTZ+3P and X-CAP-EOM-EA-CCSD/cc-PVTZ+3P levels (where X$\equiv$ box, smooth Voronoi), at fixed $\eta$ values of $0.00100\text{a.u.}, 0.00500\text{a.u.}, 0.01000\text{a.u.}$ and $ 0.02000\text{a.u.}$. 

\section{Theory}
\subsection{CAP $\eta-$trajectories and the minimum log-velocity criterion}
We start by briefly revisiting the origin\cite{riss93} of the minimum log-velocity criterion, Eq. \eqref{eq:classic_criterion}. We assume we have fixed the spatial parameters $\vb{r}^o$ and want to minimize the CAP reflection by optimizing the value of $\eta$. Our arguments below differ from the original arguments put forward in Ref. \citenum{riss93} in that we shall seek to minimize the CAP reflection by taking the unknown true Siegert energy $E_o$ as our reference, instead of the finite-basis-set-computed-energy $E(\eta)$ in the limit of $\eta \to 0$. We shall arrive at a conclusion similar to that of Ref. \citenum{riss93}, but the conceptual difference is important since the electronic state corresponding to the limit $E(\eta \to 0)$ is in most cases not a resonant state.

We begin by introducing an error function, $\epsilon(\eta)$, such that
			\begin{equation}
			\label{eq:def_error_function}
			\epsilon(\eta) := E(\eta) - E_o \ .
			\end{equation}
Naturally, the error function $\epsilon(\eta)$ is also unknown to us, and, in general, complex-valued. With a complete basis set, $E(\eta)$ will approach $E_o$ as $\eta \to 0$, i.e. $\lim_{\eta \to 0} \epsilon(\eta) = 0$. For a finite basis set, we may express $\epsilon(\eta)$ as a Maclaurin series because we expect $\eta_{opt}$ to be in the neighborhood of $\eta=0$, and Eq. \eqref{eq:def_error_function} becomes
			\begin{equation}
			\epsilon(\eta) = E(\eta) - E_o = \epsilon(0) + \sum_{n=1}\frac{\eta^n}{n!} \frac{d^n\epsilon(\eta')}{d\eta^{'n}}\Big\vert_{\eta'=0} \ .
			\end{equation}
The constant $\epsilon(0)=E(\eta=0) - E_o$ is also unknown. In first-order approximation, the absolute error, $\norm{\epsilon}$, has the expression
			\begin{equation}
			\begin{split}
			\norm{\epsilon(\eta)} = \norm{E(\eta) - E_o} & = \norm{\epsilon(0) + \eta \frac{d\epsilon(\eta')}{d\eta'}\Big\vert_{\eta'=0} + \mathcal{O}(\eta^2)} \\
			& \leq \norm{\epsilon(0)} + \norm{ \eta \frac{d\epsilon(\eta')}{d\eta'}\Big\vert_{\eta'=0}}\  \\
			& + \norm{\mathcal{O}(\eta^2)} \ ,
			\end{split}
			\end{equation}			
where we have applied the triangle inequality in the last step. Since $ \norm{\epsilon(0)}$ is unknown, we may minimize $\norm{\epsilon(\eta)}$ by minimizing $\norm{ \eta \frac{d\epsilon(\eta)}{d\eta}\Big\vert_{\eta=0}}= \norm{ \eta \frac{dE(\eta)}{d\eta}\Big\vert_{\eta=0}}$. If we further introduce the approximation 
			\begin{equation}
			\eta\frac{dE(\eta')}{d\eta'}\Big\vert_{\eta'=0} \approx \eta \frac{dE(\eta')}{d\eta'}\Big\vert_{\eta'=\eta}  
			\end{equation}
then we may consider  $\norm{\epsilon(\eta)}$ as minimized if we find an $\eta =\eta_{opt}$ such that Eq. \eqref{eq:classic_criterion} is satisfied.

Still on the first-order approximation of the error $\epsilon(\eta)$, we see that a more accurate finite-basis-set-calculated Siegert energy would be $\widetilde{E}(\eta)$, where
			\begin{subequations}
			\begin{align}
			\widetilde{E}(\eta) & \equiv E(\eta) - \eta \frac{dE(\eta')}{d\eta'}\Big\vert_{\eta'=\eta} \\
			& \approx E_o + \epsilon(0) + \mathcal{O}(\eta^2) \ .
			\end{align}
			\end{subequations}
$\widetilde{E}(\eta)$ is commonly referred to as the \textit{first-order deperturbed (Siegert) energy}. Higher-order deperturbed energies may also be calculated\cite{riss93, dempwolff22}.

\subsection{Introducing the $\xi-$criterion.}
Hitherto, practitioners of the CAP method have focused on $\eta$ when minimizing CAP reflections, especially when it comes to box-CAPs. But the operator $W$, independent of which CAP model is invoked, depends on some spatial parameters, $\vb{r}^o$. As we shall see in the next section, it is often much easier to minimize the CAP reflections through the spatial parameters $\vb{r}^o$ than through the conventional $\eta-$trajectories. 

We propose that, in general, all CAP parameters (i.e., the set $\{\eta, \vb{r}^o\}$) be amenable to optimization when minimizing CAP reflections. This also calls for an easy and applicable way of evaluating the CAP reflections -- independent of which CAP parameter we are varying to minimize CAP reflections. 

To this end, we propose an error evaluation function $\xi$ which evaluates how perturbative the CAP contribution (i.e., $-i\eta W$) is with respect to the physical electronic Hamiltonian $H$, Eq. \eqref{eq:CAP_Hamiltonian}.

Let $\cket{\psi_n}$ be a left eigenfuntion of $H_{CAP}$, the latter given by Eq. \eqref{eq:CAP_Hamiltonian}, where $H$ is the electronic Hamiltonian of a temporary anion with $N+1$ electrons. Then,
			\begin{equation}
			\label{eq:complex_eigenvalue_problem_1}
			H_{CAP} \cket{\psi_n} = E_n \cket{\psi_n}
			\end{equation}
where $E_n$ is the corresponding complex eigenenergy. 

Suppose we have solved the complex eigenvalue problem in Eq. \eqref{eq:complex_eigenvalue_problem_1} for $\cket{\psi_n}$ and $E_n$. Our aim now is to verify if the CAP, $-i\eta W$, Eq. \eqref{eq:CAP_Hamiltonian}, is truly a small perturbation on $H$, and also quantify this perturbation. 

If $E^o_{N}$ is the energy of the neutral $N$-electron system (with the nuclear positions still kept invariant), then, from \eqref{eq:complex_eigenvalue_problem_1}, we may write
			\begin{subequations}
			\begin{align}
			\cmatrixel{\psi_n}{H-E^o_N - i\eta W}{\psi_n} & = \Delta E_n\\
			\cmatrixel{\psi_n}{H-E^o_N}{\psi_n} - i \eta \cmatrixel{\psi_n}{ W}{\psi_n} & = \Delta E_n
			\end{align}		
			\end{subequations}
where $\Delta E_n \equiv E_n - E^o_N$. The following decomposition follows:
			\begin{subequations}
			\begin{align}
			\Re \Delta E_n & = \Re \cmatrixel{\psi_n}{H-E^o_N}{\psi_n} + \Re \left< - i \eta W\right>	\\
			\Im E_n=\Im \Delta E_n & = \Im \cmatrixel{\psi_n}{H}{\psi_n} + \Im \left< - i \eta W\right> 
			\end{align}
			\end{subequations}
where $\left< - i \eta W\right>_n \equiv \cmatrixel{\psi_n}{-i\eta W}{\psi_n}$. 

Note that $\Re \Delta E_n$ is the (adiabatic) vertical attachment energy (VAE) to the neutral, and $\Gamma_n \equiv -2 \Im  E_n$ is the resonance width of the temporary anion.

If the CAP, $-i\eta W$, is truly a small perturbation on $H$, then we expect
			\begin{equation}
			\label{eq:xi_Re}
			\abs{\frac{\Re \left< - i \eta W\right>_n}{\Re \Delta E_n }} \ll 1 
			\end{equation}
so that we can rule out the calculated VAE, $\Re \Delta E_n$, being a spurious effect not due to the resonance nature of the electronic state, but an artifact of the CAP. Likewise, we should also expect the imaginary part of the expectation value of the CAP, $\Im \left< - i \eta W\right>_n$, to be negligible compared to imaginary part of the total energy, $\Im E_n$, i.e.,
			\begin{equation}
			\label{eq:xi_Im}
			\abs{\frac{\Im \left< - i \eta W\right>_n}{\Im E_n }} \ll 1 \ .
			\end{equation}
In this way, we can be sure the calculated width, $\Gamma_n$, is genuine -- with little to no contribution from the CAP. 
 Hence, we may view $\frac{\Re \left< - i \eta W\right>_n}{\Re \Delta E_n }$ and $\frac{\Im \left< - i \eta W\right>_n}{\Im E_n }$ as functions quantifying CAP-reflection-errors on the VAE and $\Gamma$, respectively. Specifically, the closer the two ratios in Eq.s \eqref{eq:xi_Re} and \eqref{eq:xi_Im} are to zero, the better. Ideally (with $\eta \neq 0$ and $W \neq 0$), they should be identically zero.

We may create a single CAP error quantifying function, $\xi$, from the ratios in Eq.s \eqref{eq:xi_Re} and \eqref{eq:xi_Im} as follows:
			\begin{equation}
			\label{eq:def_xi}
			\xi \equiv \sqrt{\abs{\frac{\Re \left< - i \eta W\right>_n}{\Re \Delta E_n }}^2 + \abs{\frac{\Im \left< - i \eta W\right>_n}{\Im E_n }}^2} \ .
			\end{equation}			 
Here too, the closer $\xi$ is to zero, the better. Usually, $\abs{\frac{\Re \left< - i \eta W\right>_n}{\Re \Delta E_n }} \ll \abs{\frac{\Im \left< - i \eta W\right>_n}{\Im E_n }}$, reducing Eq. \eqref{eq:def_xi} to $\xi \approx \abs{\frac{\Im \left< - i \eta W\right>_n}{\Im E_n }}$. There are, however, regions of the CAP parameter space where the contribution of $\abs{\frac{\Re \left< - i \eta W\right>_n}{\Re \Delta E_n }}$ to $\xi$ is important, or even dominant.

Eq. \eqref{eq:def_xi} can be used to identify optimal CAP parameters. For example, say we choose to minimize the CAP reflection by varying the spatial parameters $\vb{r}^o$. Imagine $\mathcal{D}$ is the set of all the values of $\vb{r}^o$ we wish to consider. Then, on the basis of $\xi$, the optimal $\vb{r}^o$ belonging to $\mathcal{D}$, $\vb{r}^o_{opt}$, is the $\vb{r}^o \in \mathcal{D}$ which minimizes $\xi$. That is,
			\begin{equation}
			\vb{r}^o_{opt} = \min_{\mathcal{D}}\ \xi(\vb{r}^o).
			\end{equation}
In general terms, 
			\begin{equation}
			\label{eq:xi_criterion}
			X_{opt} = \min_{X \in \mathcal{D}} \ \xi(X) \ .
			\end{equation}
where $X$ can be any CAP parameter (or combination of parameters) and $\mathcal{D}$ is a given set of values of $X$. Eq. \eqref{eq:xi_criterion} constitutes a general criterion for choosing an optimal value for any CAP parameter $X$. We refer to Eq. \eqref{eq:xi_criterion} as the $\xi-$criterion. Empirically, we can say that the CAP reflections on a computed resonant state is acceptable when the $\xi$ value is $\approx 0.05$, and even better if $\xi<0.05$; computed resonant states with $\xi\lesssim 0.01$ should be the gold-standard.

We also note that as long as $\left<-i \eta W \right>_n$ can be computed, we can also compute $\xi$, Eq. \eqref{eq:def_xi}. The computation of $\xi$ can always be done, independent of which CAP parameter(s) we are trying to optimize. Thus, unlike the minimum log-velocity criterion, Eq. \eqref{eq:classic_criterion}, which is specific to $\eta-$trajectories, the new criterion in Eq. \eqref{eq:xi_criterion} is always applicable.

If $E_n=\cmatrixel{\psi_n}{H}{\psi_n} +\left< - i \eta W\right>_n$, and $\left< - i \eta W\right>_n$ is a perturbation, then the corresponding  deperturbed (or corrected) Siegert energy, $\widetilde{E}_n$, is simply
			\begin{subequations}
			\label{eq:def_new_corrected_E}
			\begin{align}
			\widetilde{E}_n & = E_n - \left< - i \eta W\right>_n \label{eq:def_new_corrected_E_a}\\
			\widetilde{E}_n & = \cmatrixel{\psi_n}{H}{\psi_n} \label{eq:def_new_corrected_E_b}\ .
			\end{align}
			\end{subequations}
Thus, the deperturbed Siegert energy $\widetilde{E}_n$ is nothing but the expectation value of the physical electronic Hamiltonian $H$ with respect to the eigenfunction $\cket{\psi_n}$ of $H_{CAP}$. We point out that it would be amiss to refer to $\widetilde{E}_n$ as a 'first-order corrected energy' since, as can be inferred from Eq. \eqref{eq:def_new_corrected_E_b}, the dependence of  $\widetilde{E}_n$ on CAP parameter $X$ stems from $\cket{\psi_n}$ -- therefore, in a nonlinear way. Indeed, it is possible to expand $\widetilde{E}_n$ in powers of any CAP parameter.

\subsubsection{The complex self-consistent field (CSCF) procedure as a dynamical system.}
When minimizing CAP reflections by optimization of the spatial parameters of a box- or smooth Voronoi- CAP at the level of Hartree-Fock (HF), we have observed that it is not uncommon to veer into regions of these parameter space with bifurcations (to be explained soon).
To fully understand them, we need the theory of nonlinear dynamics\cite{strogatz18,thompson02} which is beyond the scope of the present paper, but we will briefly discuss the pertinent salient notions of the theory here.

As is well-known, the HF self-consistent field (SCF) method is an iterative method for determining the eigenvectors and eigenvalues of a given electronic Hamiltonian.  A similar SCF procedure is employed for non-Hermitian HF problems in general\cite{whiteNH15,theel17}, which is usually referred to as complex SCF (CSCF)\cite{whiteNH15}. Since the CSCF is iterative, we may liken it to a discrete dynamical system\cite{theel17}. In this sense, each iteration step of the CSCF may be viewed as a time-step of the dynamics. Given a convergence criterion, the final solution of the CSCF procedure may be viewed as a \textit{stable attractor}. A stable attractor is a state toward which the dynamical system evolves to, starting from a wide range of initial conditions. All initial conditions of a dynamical system which evolve towards the same stable attractor are referred to as the attractor's \textit{basin of attraction}. In the case of the CSCF procedure as a dynamical system, the guessed electronic density matrix is the initial condition.

The equations governing the evolution of a dynamic system are also dependent on a set of parameters.
In the study of dynamic systems, these parameters are important as they can dramatically change the qualitative nature of the dynamics, e.g., the disappearance and appearance of stable attractors. In viewing the CAP-HF CSCF procedure as a dynamic system, the CAP parameters become, naturally, the parameters of the dynamics. It is usually the case that a basin of attraction of the CAP-HF CSCF procedure has a single stable attractor (or solution), the latter being dependent on the value of the CAP parameters. However, as we shall show below, as we change the CAP spatial parameters $\vb{r}^o$ (with $\eta$ fixed), the CSCF as a dynamical system may have, in a certain region of $\vb{r}^o$, more than one stable attractor. In such a case, with the same guess for the electronic density matrix and CAP parameters, the repetition of the CAP-HF CSCF procedure will not produce the same solution but will randomly settle on one of a finite set solutions. In nonlinear dynamics theory\cite{strogatz18,thompson02}, this phenomenon whereby a basin of attraction have more than one attractor is referred to as \textit{bifurcation}. Bifurcations have previously been reported in Hartree-Fock calculations\cite{theel17,natiello84}.

There are several types of bifurcation; the interested reader may consult the literature\cite{strogatz18,thompson02}. Among the various types of bifurcation, the relevant one to the present paper is the so-called \textit{pitchfork bifurcation}, which is the type of bifurcation we have observed for the CAP-HF CSCF procedure. Here, the bifurcation leads to three attractors, two of which form a symmetrical pair. For real-valued fixed point attractors, this symmetrical pair character would mean two attractors are opposite of each other; while for complex-valued fixed point attractors, it would mean two attractors are the complex conjugate of each other.

In concrete terms, in the event of a pitchfork bifurcation in the CAP-HF CSCF procedure,  upon repetition of the same static energy calculation (with the same initial guess for the electronic density matrix and CAP parameters), one observes three possible states: $S_+, S_-$ and $S_0$, characterized by complex eigenenergies $E_1, E_2$ and $E_3$, respectively -- with $E_1^* = E_2$, and $\Im E_3 \sim 0$ (naturally, $\Re E_1 \neq 0$, and $\Re E_3 \neq 0$), where $x^*$ indicates complex conjugate of $x$. One important feature of the attractor states $S_+$ and $S_-$ is that the magnitude of the CAP reflections on their energies $E_1$ and $E_2$ is extremely low. In fact, the asymptotic limits of $S_+$ and $S_-$ have 
			\begin{equation}
			\Re \left<-i \eta W \right> = \Im \left<-i \eta W \right> = 0 \ .
			\end{equation}
despite the nonzero values of the complex conjugate pair $E_1, E_2$ and the nonzero CAP parameters $\{\eta, \vb{r}^o\}$. The asymptotic limit of the $S_0$ state is the solution of the CAP-HF CSCF when $\eta=0$. The $S_+$ state, which we define as having a negative imaginary part of energy (i.e., $\Im E_1 <0$) represents a decay resonant state, while the $S_-$ state (with $\Im E_2 >0$) represents a capture resonant state\cite{moiseyev11}. 

Our computations show that the advent of a pitchfork bifurcation for CAP-HF CSCF is driven by an interval of relatively large values of the spatial CAP parameters $\vb{r}^o$. Beyond this interval, the pitchfork bifurcation disappears and $S_0$ becomes the sole attractor.

It is important to note that, in general, if a decay resonant state is the sole stable attractor for a given $H_{CAP}$, then its corresponding capture resonant state is the corresponding eigenstate of $H_{CAP}$'s adjoint, $H^\dagger_{CAP}$, i.e.,
			\begin{equation}
			H^\dagger_{CAP} = H + i \eta W \ .
			\end{equation}
It is worth noting that changing the sign of $\eta$ in $H_{CAP}$ is equivalent to taking the adjoint of $H_{CAP}$, and vice versa:
			\begin{equation}
			\lim_{\eta \to -\eta} H_{CAP} = H^\dagger_{CAP} \ .
			\end{equation}
From this point of view, each of the three attractors appearing in a pitchfork bifurcation discussed above may be associated with three distinct values of $\eta$: 
			\begin{equation}
			\begin{split}
			S_+ & \leftrightarrow +\eta \\ 
			S_- & \leftrightarrow - \eta \\
			S_0 & \leftrightarrow \eta=0 \ .
			\end{split}
			\end{equation}
Given that in the analyses we embark on in the next section, we hold $\eta$ constant and vary only the spatial CAP parameters $\vb{r}^o$, it is remarkable that the appearance of pitchfork bifurcations in the CAP-HF calculations are related to distinct values of $\eta$.

For a fixed $\eta$, there may be a collection of $\vb{r}^o$ values where we observe a pitchfork bifurcation in the CAP-HF calculations.  We may refer to such a collection  as the \textit{pitchforck bifurcation region} (PBR).

\section{Results and discussion}
In this section, we analyze and optimize CAP-HF and CAP-EOM-EA-CCSD solutions for the temporary anion of dinitrogen. For each anion, we shall determine the resonance energy and width by optimizing the box- and smooth Voronoi- CAPs at the CAP-HF and CAP-EOM-EA-CCSD levels.

We emphasize here that for the box-CAP, one can minimize the CAP reflections through an arbitrary variation scheme of the onsets $\{r^o_\alpha\}$. There is an infinite number of such schemes. For example, one may fix, besides $\eta$, $r^o_x, r^o_y$ and vary $r^o_z$. Or, one could also fix $r^o_x$ and $r^o_z$, and vary $r^o_y$ according to the relation $r^o_y = (r^o_x+r^o_z)\cdot c^{-n}$ with $n=0,1,2,3, \ldots$ and $c$ a positive real constant. It is advisable, however, to choose a variation scheme which does not break the point symmetry of resonant states.

In this paper, we shall use the simple scheme whereby the box-CAP onset is the same in all directions, i.e., $r^o_x = r^o_y = r^o_z = r^o$, where $r^o$ is our variation parameter. The advantage of this choice is that it makes the comparison between results from box- and smooth Voronoi-CAPs easy.

%To avoid repeating detailed discussions of the same observations, in the coming sections, we shall discuss the results of $r^o-$trajectories for \ce{N2-} (\S \ref{subsec:dinitrogen}) in much details compared to \ce{C2H4-} (\S \ref{subsec:ethylene}) and cis-\ce{HCOOH-} (\S \ref{subsec:formic}). 

\subsection{Dinitrogen}
\label{subsec:dinitrogen}
\subsubsection{Box-CAP $r^o_\alpha$ trajectory analysis}
\begin{figure*}
\includegraphics[scale=0.50]{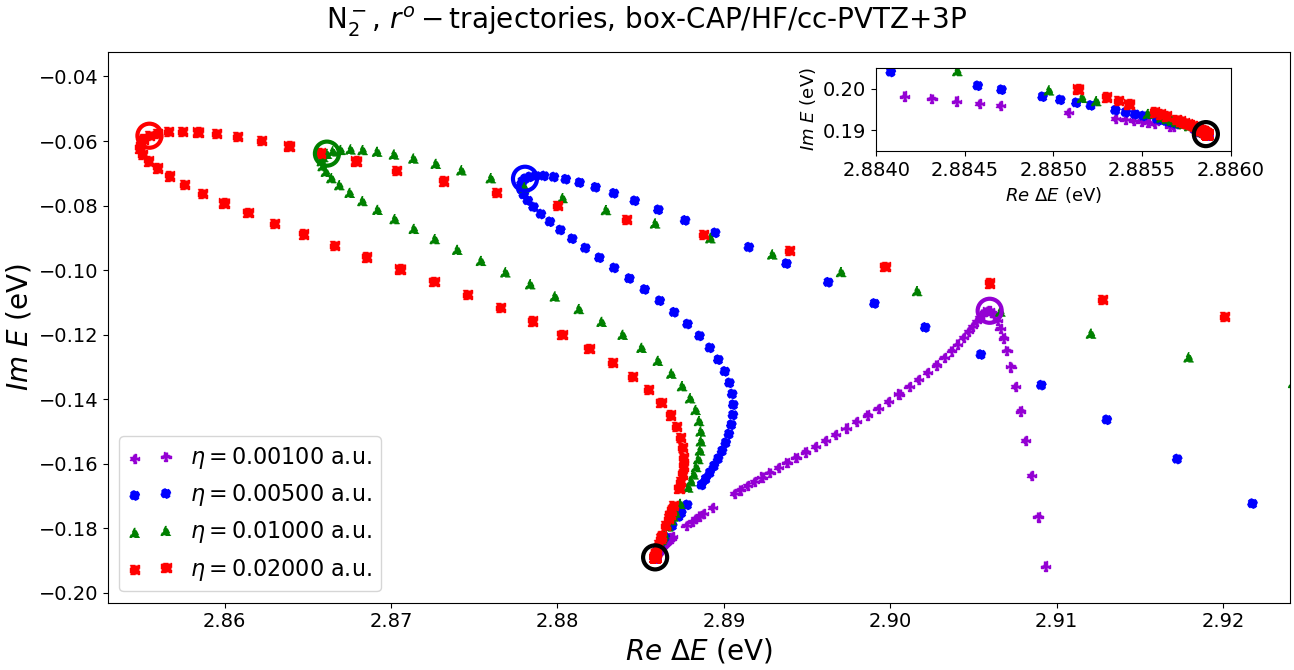}
\caption{Snippet of $r^o-$trajectories for the $^2\Pi_{g}$ resonance of \ce{N2-} computed at the box-CAP/HF/cc-PVTZ+3P level. Colored rings are put around points with minimum $\xi$. Each $r^o_-$trajectory was computed at steps of $\Delta r^o = 0.100a_o$.}
 \label{fgr:gg1}
\end{figure*}

\begin{figure*}
\includegraphics[scale=0.50]{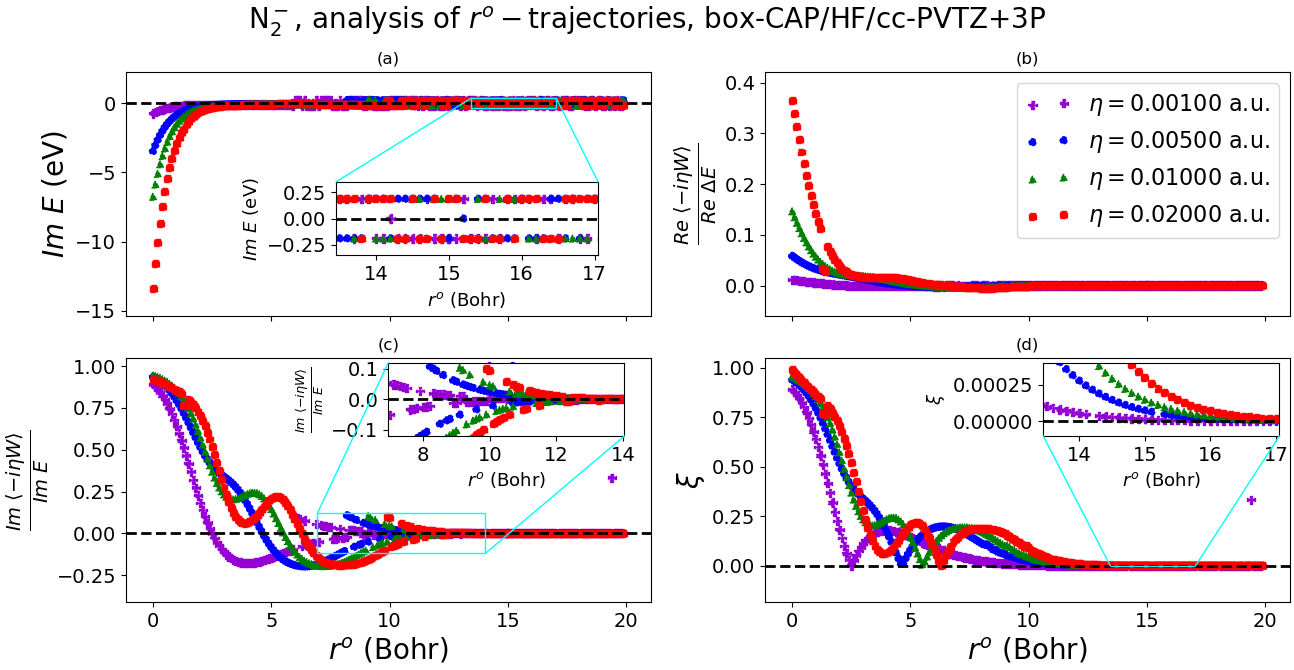}
\caption{Analysis of $r^o-$trajectories for the $^2\Pi_{g}$ resonance of \ce{N2-} computed at the box-CAP/HF/cc-PVTZ+3P level.}
 \label{fgr:gg2}
\end{figure*}

\paragraph{\textbf{Box-CAP/HF, \ce{N2-}}}
In Fig. \ref{fgr:gg1} we show snippets of the distribution of the eigenvalues, in the complex plane, of the box-CAP Hamiltonian for the \ce{N2-} computed at the HF level. Each distribution was computed at a fixed $\eta$ value. The results of the analysis of the distributions are shown in Fig. \ref{fgr:gg2}. 

If we look at the $\xi$ curves in panel (d) of Fig. \ref{fgr:gg2}, we note that each curve presents a minimum at a certain $r^o$ which is below $7.5a_o$ ($1a_o \equiv$ 1 Bohr), and approaches the line $\xi=0$ for $r^o >11a_o$. As mentioned above, we are interested in the minima of $\xi$, since these indicate optimal CAP parameters according to the discussions above. These minima may or not correspond to the same resonant state. 

Despite having different widths (see Tab. \ref{tbl:dinitrogen}), careful scrutiny of the absolute minima of $\xi$ at $r^o< 7.5a_o$ suggests that they all correspond to the same resonant state but are characterized by different momentum of the extra electron. The complex energies corresponding to these absolute minima are encircled in violet, blue, green and red rings for the curves $\eta=0.000100$a.u., $0.000500$a.u., $0.001000$a.u., and $0.002000$a.u., respectively, in Fig. \ref{fgr:gg1}; the fact that the points move on the complex plane as we change $\eta$ may indicate there is still some coupling between the resonant state and the continuum. We also note that the same encircled points in question are always embedded in a region of complex plane with a high density of points.

From Fig. \ref{fgr:gg1}, we also observe the presence of two regions with very high density of points, their center of attraction encircled in black rings. Those points originate from the pitchfork bifurcation. The centers and the neighboring points represent the decay (when $\Im E <0$) and capture (when $\Im E >0$) resonant states when pitchfork bifurcation occurs. The centers are complex conjugate of each other and are the asymptotic limits of their respective type of resonant state. 

A crucial characteristic of these center of attractions is the fact that: i) they are invariant with respect to $\eta$, and ii) the real and imaginary parts of the expectation value for the CAP, $\left< -i \eta W\right>$, are identically zero at these points (thus, $\xi=0$). In addition, we may note that the computed VAE and width at the center of attraction for decay resonance of $^2\Pi_g$ \ce{N2-} is very close to the experimental value and agrees also well with results from CAP-EOM-EA-CCSD calculations (see Tab. \ref{tbl:dinitrogen}).

The pitchfork bifurcation region (PBR) is clearly evident in the $\frac{\Im \left<-i\eta W \right>}{\Im E}$ versus $r^o$ plot, i.e., Fig. \ref{fgr:gg2} panel (c), where, due to the presence of complex conjugated pair of eigenenergies, the points mimic a reflection across the $\frac{\Im \left<-i\eta W \right>}{\Im E}=0$ line (i.e., the dark dashed line in the figure). Part of the PBR is shown in the inset of Fig. \ref{fgr:gg2} panel (c). Fig. \ref{fgr:gg2} panel (c) also clearly shows that as we increase $r^o$ from $r^o=0$, there is a critical value of $r^o$, $r^o_{critical}$, beyond which we see a bifurcation of the box-CAP/HF solutions. The value of $r^o_{critical}$ depends on the fixed value of $\eta$: the higher $\eta$, the higher $r^o_{critical}$. For example, for the $^2\Pi_g$ \ce{N2-} resonance, $r^o_{critical} \sim 6a_o$ for $\eta=0.00100$ a.u., $r^o_{critical} \sim 8.2a_o$ for $\eta=0.00500$ a.u., and $r^o_{critical} \sim 9a_o$ for $\eta=0.01000$ a.u. (see Fig. \ref{fgr:gg2}).

The conjugate pair of attractors in the PBR approach their respective asymptotic limits as $r^o$ increases. The asymptotic limits also form a conjugate pair  and have no CAP reflections (thus, $\Re \left<-i\eta W \right> = \Im \left<-i\eta W \right> = 0$), and consequently, $\xi=0$. This explains why we see points in the PBR region in Fig. \ref{fgr:gg2}, panel (c), approaching the $\frac{\Im \left<-i\eta W \right>}{\Im E}=0$ line (black dashed line in the figure) as $r^o$  increases. In the $\xi$ versus $r^o$ plot, we also see a similar trend. 

It must, however, be emphasized that for $r^o$ beyond the PBR, we still find $\Re \left<-i\eta W \right> = \Im \left<-i\eta W \right> = 0$, but this time, $Im E =0$ (thus, $\xi$ is not defined) and the CAP-HF solutions coincides with the $\eta=0$ solution.

An even more reliable and easy way to spot a PBR is by looking at the $\Im E$ versus $r^o$ plot, where in the PBR -- besides points lying on (or very close) to the  $\Im E=0$ line -- one will see two series of points, each series on a horizontal line and the horizontal lines will be equidistant from the $\Im E=0$ line (in other words, the series of points will be opposite each other). This behavior is shown in the inset of Fig. \ref{fgr:gg2}, panel (a).  

\begin{table*}
\begin{ruledtabular}
\begin{tabular}{lllclllll}
% \hline
 & $\eta$ (a.u.) & $r^o_{opt}$ (Bohr) & $\xi(r^o_{opt}) $ & 
VAE (eV)  & VA$\widetilde{\text{E}}$ (eV) & $\Gamma$ (eV) & $\widetilde{\Gamma}$ (eV) \\ 
\hline
%\hline
Box-CAP/HF  &		  & 	   &			&			 &			 &			 &			\\
 			& 0.00100 &	2.500  &  $0.00055$	 & 	2.9060	 & 	2.9075	 & 	0.2252	 & 	0.2252	\\
 			& 0.00500 &	4.600  &  $0.01002$	 & 	2.8780 	 & 	2.8763	 & 	0.1434	 & 	0.1419	\\
 			& 0.01000 &	5.500  &  $0.00644$	 & 	2.8661	 & 	2.8641	 & 	0.1278	 & 	0.1270	\\
 			& 0.02000 &	6.300  &  $0.00207$	 & 	2.8555	 & 	2.8521	 & 	0.1167	 & 	0.1169	\\
 			& all 	  &	PBR    &  $0.00000$	 & 	2.8859	 & 	2.8859	 & 	0.3781	 & 	0.3781	\\
Smooth Voronoi-CAP/HF  &		  & 	   &			&			 &			 &			 &			\\
 			& 0.00100 &	2.900  &  $0.01021$	 & 	2.9061	 & 	2.9065	 & 	0.2173	 & 	0.2151	\\
 			& 0.00500 &	5.700  &  $0.00262$	 & 	2.8752 	 & 	2.8741	 & 	0.1342	 & 	0.1345	\\
 			& 0.01000 &	6.700  &  $0.00528$	 & 	2.8620	 & 	2.8608	 & 	0.1157	 & 	0.1151	\\
% 			& 0.02000 &	4.600  &  $0.01525$	 & 	2.9563	 & 	2.9116	 & 	0.2466	 & 	0.2461	\\
 			& 0.02000 &	7.600  &  $0.00189$	 & 	2.8497	 & 	2.8486	 & 	0.1020	 & 	0.1018	\\
 			& all 	  &	PBR    &  $0.00000$	 & 	2.8859	 & 	2.8859	 & 	0.3781	 & 	0.3781	\\
Box-CAP/EOM-EA-CCSD   &		   & 	         &			 &			 &			 &			 &			\\
 			& 0.00100 &	2.400  &  $0.43036$	 & 	2.7673	 & 	2.8497	 & 	0.3678	 & 	0.2099	\\
% 			& 0.00500 &	2.400  &  $0.35587$	 & 	2.8578	 & 	2.6438	 & 	0.7373	 & 	0.4808	\\
 			& 0.00500 &	4.500  &  $0.00984$	 & 	2.6724	 & 	2.6987	 & 	0.3057	 & 	0.3057	\\
% 			& 0.00500 &	5.900  &  $0.06431$	 & 	2.7116	 & 	2.8852	 & 	0.3337	 & 	0.3357	\\
% 			& 0.01000 &	3.100  &  $0.07976$	 & 	2.8944	 & 	2.6929	 & 	0.5902	 & 	0.5673	\\
 			& 0.01000 &	5.300  &  $0.00894$	 & 	2.6421	 & 	2.6473	 & 	0.2729	 & 	0.2705	\\
%  			& 0.01000 &	7.100  &  $0.08334$	 & 	2.7042	 & 	2.9195	 & 	0.3663	 & 	0.3573	\\
%  			& 0.02000 &	3.300  &  $0.04342$	 & 	2.9789	 & 	2.8595	 & 	0.5399	 & 	0.5489	\\
%  			& 0.02000 &	4.300  &  $0.07275$	 & 	2.8210	 & 	2.6204	 & 	0.4556	 & 	0.4627	\\
  			& 0.02000 &	6.100  &  $0.00797$	 & 	2.6164	 & 	2.6127	 & 	0.2375	 & 	0.2394	\\
%  			& 0.02000 &	8.100  &  $0.10314$	 &  2.6966	 & 	2.9477	 & 	0.3988	 & 	0.3811	\\

Smooth Voronoi-CAP/EOM-EA-CCSD   &		   & 	         &			 &			 &			 &			 &			\\ 
 			& 0.00100 &	2.900  &  $0.36639$	 & 	2.7644	 & 	2.8470	 & 	0.3407	 & 	0.2163	\\
% 			& 0.00500 &	2.800  &  $0.21514$	 & 	2.8654	 & 	2.6546	 & 	0.6208	 & 	0.4953	\\
 			& 0.00500 &	5.400  &  $0.00586$	 & 	2.6621	 & 	2.6770	 & 	0.2806	 & 	0.2811	\\
% 			& 0.00500 &	7.500  &  $0.08333$  &  2.7029   &  2.9054   &  0.3472   &  0.3599	\\
% 			& 0.01000 &	3.000  &  $0.05498$  &  2.9763   &  2.8244   &  0.6054   &  0.5930	\\
%			& 0.01000 & 4.200  &  $0.07037$  &  2.8272   &  2.6319   &  0.4579   &  0.4517	\\
 			& 0.01000 &	6.400  &  $0.01744$	 & 	2.6267	 & 	2.6301	 & 	0.2363	 & 	0.2404	\\
%  			& 0.01000 &	8.800  &  $0.09393$  &  2.6941   &  2.9456   &  0.3901   &  0.3860	\\
%  			& 0.02000 &	3.500  &  $0.03590$  &  3.0114   &  2.9176   &  0.4759   &  0.4674	\\
%  			& 0.02000 &	5.300  &  $0.06855&  &  2.7920   &  2.6036   &  0.4080   &  0.4031	\\
  			& 0.02000 &	7.200  &  $0.03460$	 & 	2.6011	 & 	2.5953	 & 	0.1990	 & 	0.1921	\\
%  			& 0.02000 &	9.800  &  $0.13762$  &  2.6757   &  2.9651   &  0.4235   &  0.4595	\\
Reference data &		  & 	   &			 &			 &			 &			 &			\\
  			& 		  &		   &  			 &  2.32		 & 			 & 	0.41	 & 			\\
\end{tabular} 
\caption{Summary of the analysis of the $r^o-$trajectories for the $^2\Pi_g$ resonance of \ce{N2-}. The $r^o$ values reported in the table are the optimal values according to the $\xi-$criterion, Eq. \ref{eq:xi_criterion}. For box-CAP computations, $r^o_x=r^o_y=r^o_z\equiv r^o$. VA$\widetilde{\text{E}}$ and $\widetilde{\Gamma}$ are the corrected VAE and resonance widths, respectively.}
\label{tbl:dinitrogen}
\end{ruledtabular}
\end{table*}

\begin{figure*}[!h]
\includegraphics[scale=0.50]{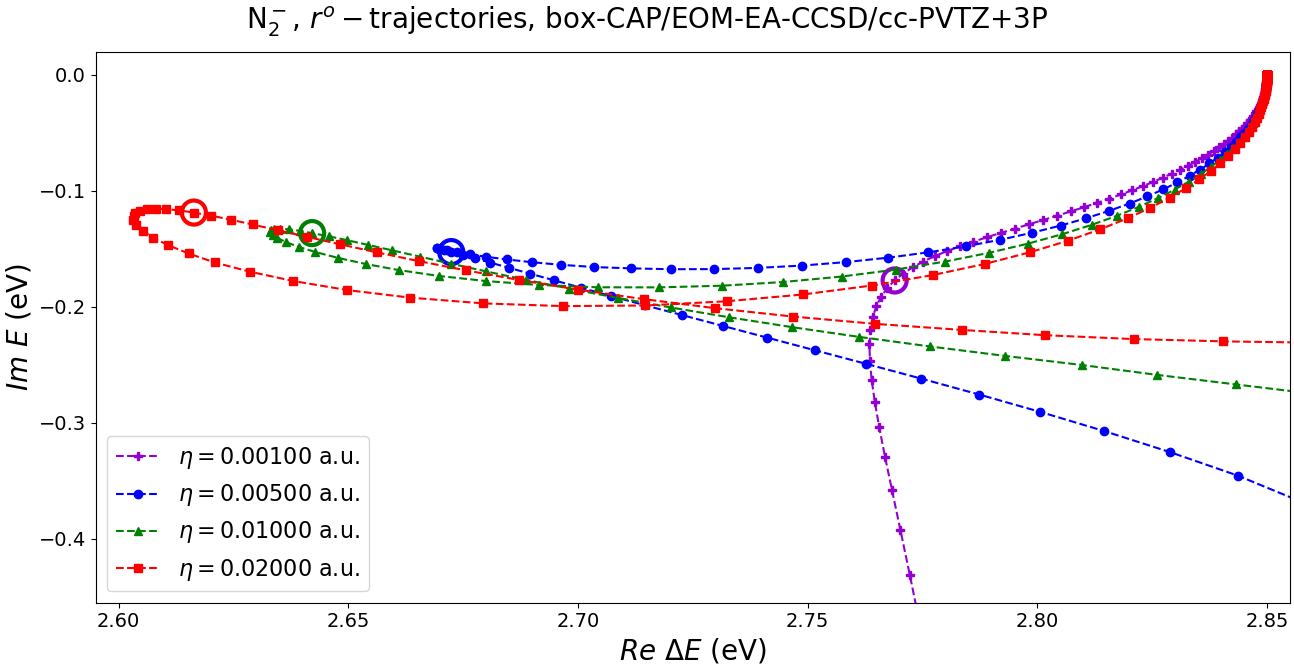}
\caption{Snippet of $r^o-$trajectories for the $^2\Pi_{g}$ resonance of \ce{N2-} computed at the box-CAP/EOM-EA-CCSD/cc-PVTZ+3P level. Colored rings are put around points with minimum $\xi$. Each $r^o_-$trajectory was computed at steps of $\Delta r^o = 0.100a_o$.}
 \label{fgr:gg3}
\end{figure*}

\paragraph{\textbf{Box-CAP/EOM-EA-CCSD, \ce{N2-}}}
The distribution of the Siegert energies (computed at the box-CAP/EOM-EA-CCSD/cc-PVTZ+3P level) of the $^2\Pi_g$  state of \ce{N2-} in the complex plane is shown in Fig. \ref{fgr:gg3} for different fixed values of $\eta$. We did not see any bifurcations at relatively large $r^o$ values like those observed at the box-CAP/HF/cc-PVTZ+3P level, Fig. \ref{fgr:gg1}.

\begin{figure*}[!h]
\includegraphics[scale=0.50]{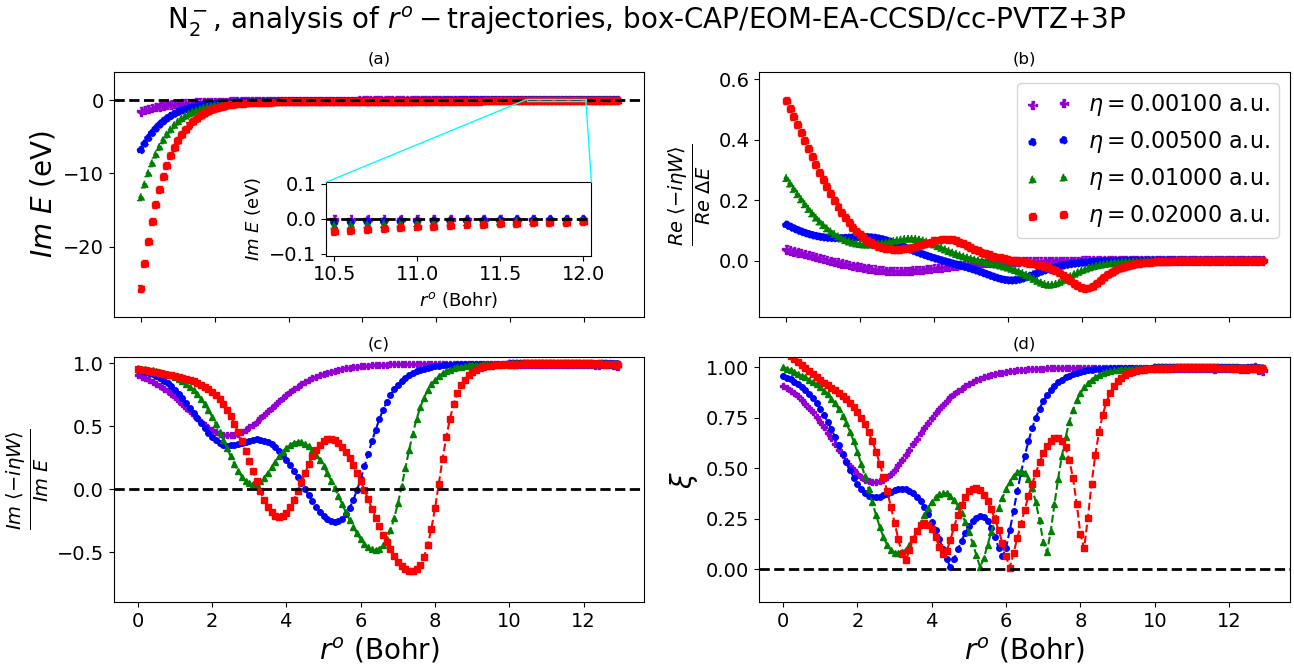}
\caption{Analysis of $r^o-$trajectories for the $^2\Pi_{g}$ resonance of \ce{N2-} computed at the box-CAP/EOM-EA-CCSD/cc-PVTZ+3P level.}
 \label{fgr:gg4}
\end{figure*}

We also note that in the $\xi$ versus $r^o$ plots in Fig. \ref{fgr:gg4}, panel (d), $\xi$ is $\sim 1$ for $r^o$ greater than $10a_o$, for all $\eta$ values considered; that indicates strong CAP reflections on the computed complex energies. Comparing panels (b) and (c) of Fig. \ref{fgr:gg4}, we notice that the ratio $\frac{\Im \left< -i\eta W\right>}{\Im E}$ tends to unity for large $r^o$ while $\frac{\Re \left< -i\eta W\right>}{\Re \Delta E} $ tends to 0. Hence, the source of the large CAP reflections is the imaginary part of computed energy, $\Im E$. $\frac{\Im \left< -i\eta W\right>}{\Im E} \to 1$ for large $r^o$ implies that the imaginary parts of the computed complex energies are not related to the physical widths of the resonant anion but are just the imaginary part of the expectation value of the CAP: thus, from Eq. \eqref{eq:def_new_corrected_E}, it follows that the imaginary part of the corresponding corrected complex energies will be practically zero.

In Table \ref{tbl:dinitrogen}, we report the values for $r^o_{opt}$ of the box-cubes with the absolute minimum $\xi$ for the four $\eta$ values considered and their related computed vertical attachment energies (VAEs) and widths (corrected and uncorrected). 

We note that of the four $\eta$ values, $\eta=0.00100$a.u. had a minimum $\xi$ of about $0.43$, which is way above the acceptable value of $0.05$. Such a high value of $\xi$ translates into a larger CAP reflection on the resonance energy. Not surprisingly, the corresponding corrected and uncorrected VAEs differ by about $0.11$eV, and so do the computed corrected and uncorrected widths. 

For the $\eta=0.00500\text{a.u.}, 0.01000\text{a.u.}, 0.02000\text{a.u.}$ curves, the absolute minimum of $\xi$ is below $0.01$, and the computed corrected and uncorrected VAEs differ by $0.03$eV or below, while the values for the corrected and uncorrected widths are even closer. For these $\eta$ curves, we also observe a good agreement between the VAEs and widths from one curve to the other.

\subsubsection{Smooth Voronoi-CAP $r^o-$trajectory analysis}
\begin{figure*}[!h]
\includegraphics[scale=0.50]{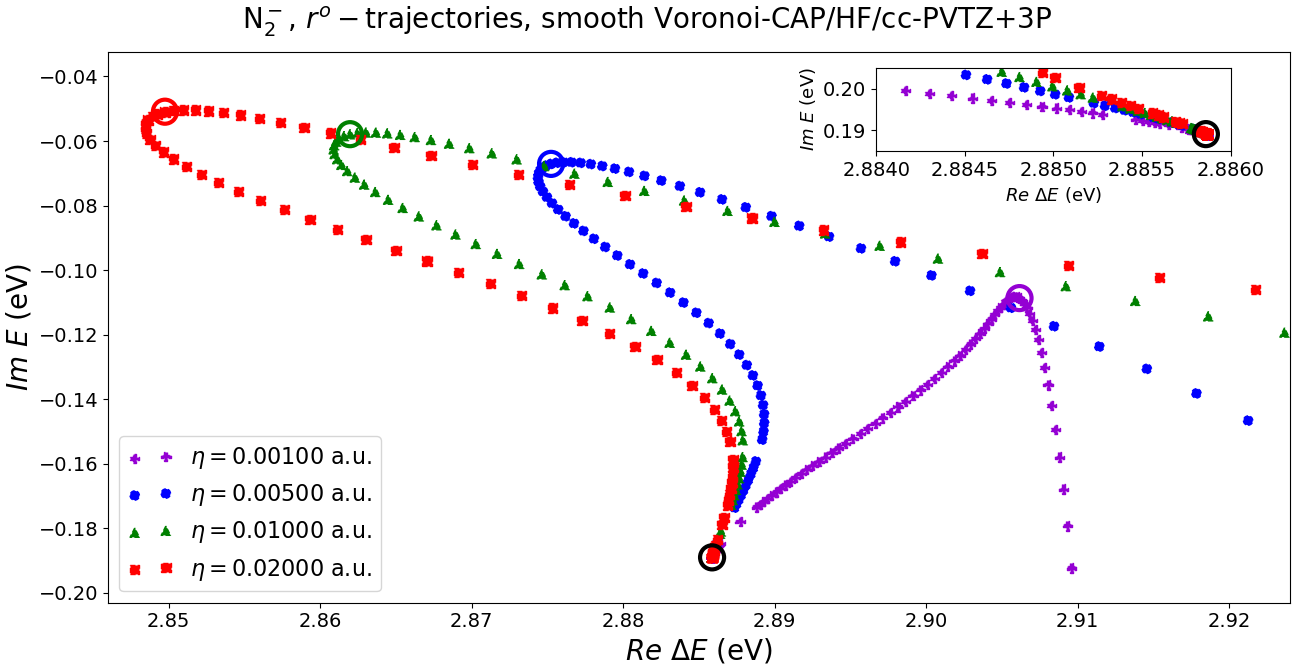}
\caption{Snippet of $r^o-$trajectories for the $^2\Pi_{g}$ resonance of \ce{N2-} computed at the smooth Voronoi-CAP/HF/cc-PVTZ+3P level. Colored rings are put around points with minimum $\xi$. Each $r^o_-$trajectory was computed at steps of $\Delta r^o = 0.100a_o$.}
 \label{fgr:gg5}
\end{figure*}

\begin{figure*}[!h]
\includegraphics[scale=0.50]{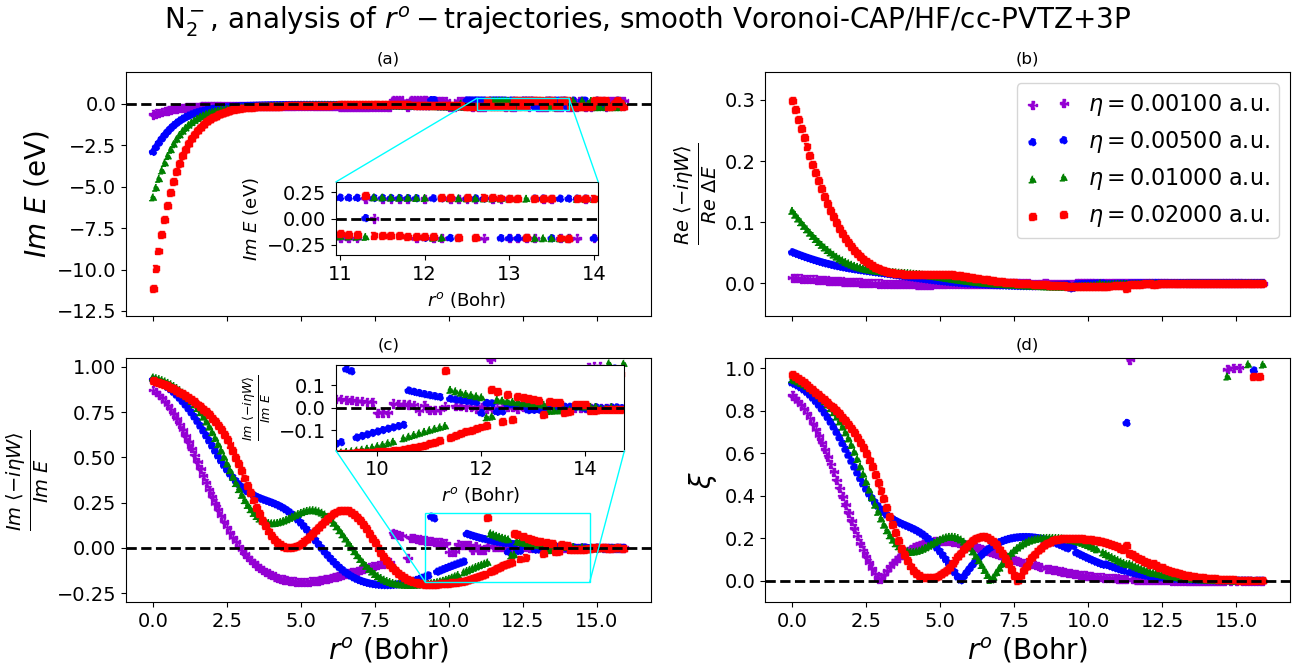}
\caption{Analysis of $r^o-$trajectories for the $^2\Pi_{g}$ resonance of \ce{N2-} computed at the smooth Voronoi-CAP/HF/cc-PVTZ+3P level.}
 \label{fgr:gg6}
\end{figure*}

\paragraph{\textbf{Smooth Voronoi-CAP/HF, \ce{N2-}}}
The results are very similar to those obtained with box-CAP/HF and are plotted in Fig.s \ref{fgr:gg5} and \ref{fgr:gg6} and summarized in Table \ref{tbl:dinitrogen}. Here too, we observe the presence of a pitchfork bifurcation region (PBR), and coincides more or less with that of the box-CAP/HF. The energy of the asymptotic limit of the conjugate pair of attractors in the PBR (also encircled by black rings in Fig. \ref{fgr:gg5}) is \textit{exactly} the same as that of the box-CAP/HF calculations (see Table \ref{tbl:dinitrogen}).

\begin{figure*}[!h]
\includegraphics[scale=0.50]{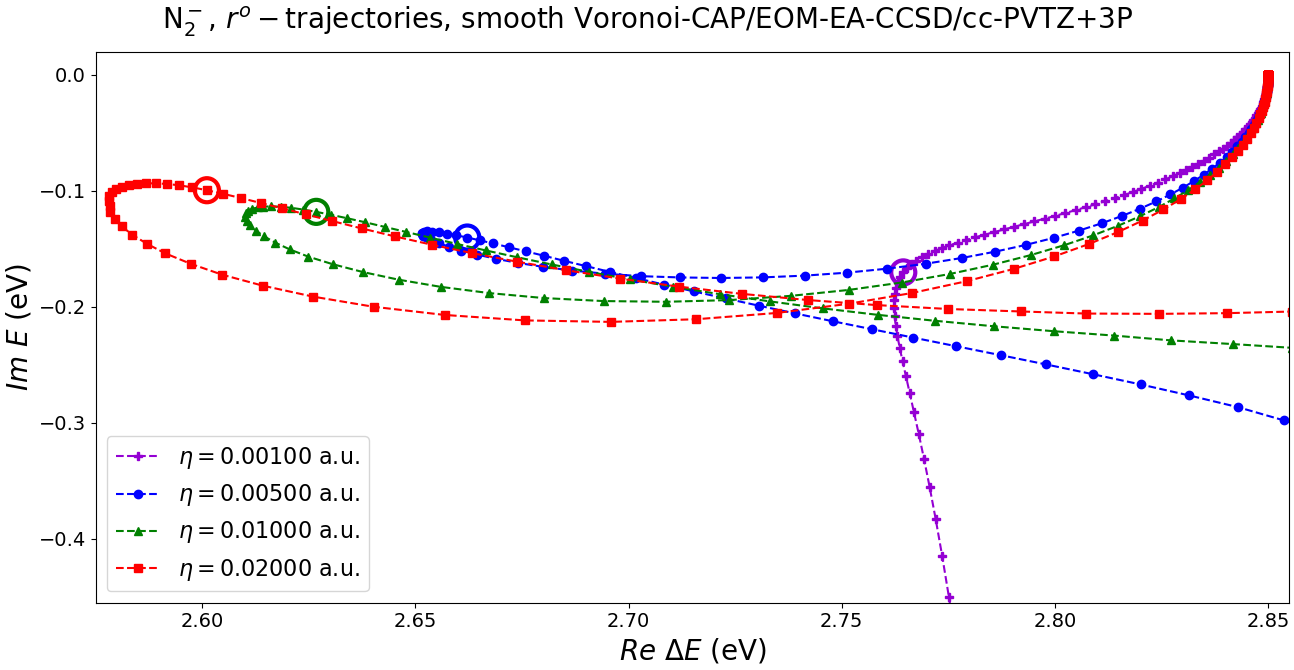}
\caption{Snippet of $r^o-$trajectories for the $^2\Pi_{g}$ resonance of \ce{N2-} computed at the smooth Voronoi-CAP/EOM-EA-CCSD/cc-PVTZ+3P level. Colored rings are put around points with minimum $\xi$. Each $r^o_-$trajectory was computed at steps of $\Delta r^o = 0.100a_o$.}
 \label{fgr:gg7}
\end{figure*}

\begin{figure*}[!h]
\includegraphics[scale=0.50]{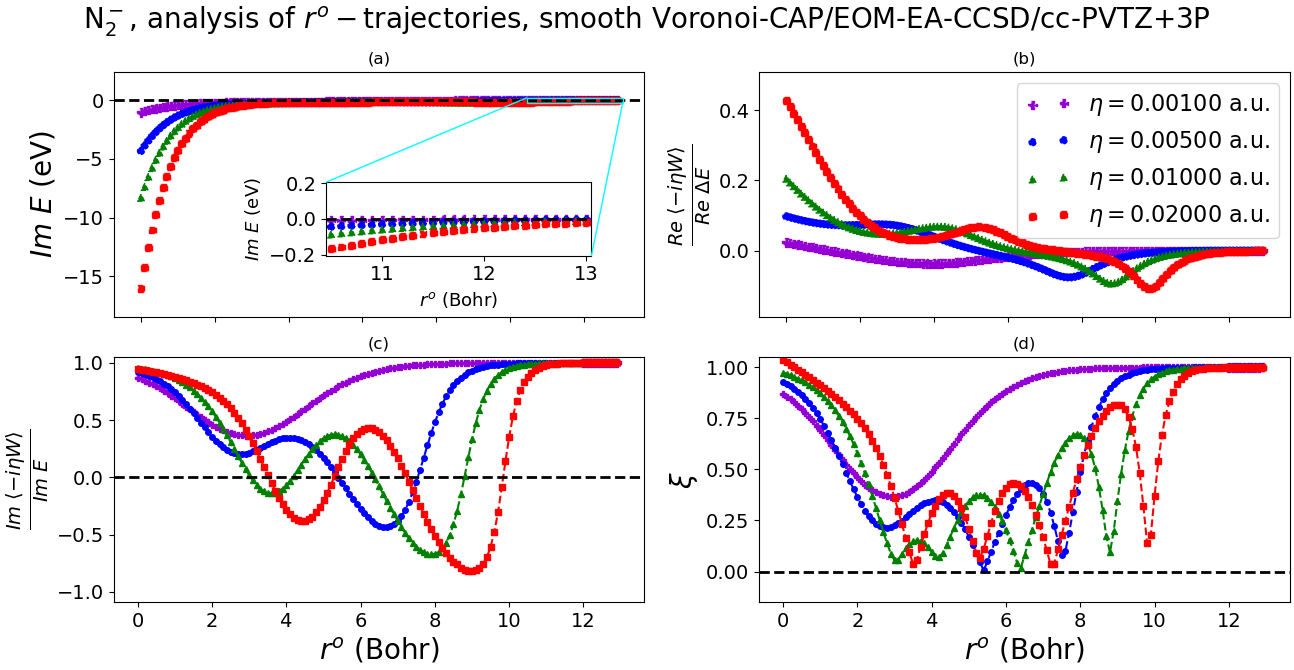}
\caption{Analysis of $r^o-$trajectories for the $^2\Pi_{g}$ resonance of \ce{N2-} computed at the smooth Voronoi-CAP/EOM-EA-CCSD/cc-PVTZ+3P level.}
 \label{fgr:gg8}
\end{figure*}

\paragraph{\textbf{Smooth Voronoi-CAP/EOM-EA-CCSD, \ce{N2-}}}
The results for smooth Voronoi-CAP/EOM-EA-CCSD (Fig.s \ref{fgr:gg7} and \ref{fgr:gg8}) also match with the box-CAP/EOM-EA-CCSD results (Fig.s \ref{fgr:gg3} and \ref{fgr:gg4}). No bifurcation points were observed. 

From Table \ref{tbl:dinitrogen}, we observe that, for a fixed $\eta$, the optimal CAP box dimension according to the $\xi-$criterion, is approximately the same between HF and EOM-EA-CCSD; similar observation  holds for smooth Voronoi-CAP/HF and smooth Voronoi-CAP/EOM-EA-CCSD. In addition, at the same level of theory, the optimal $r^o$ is approximately the same, for fixed $\eta$, whether we use box-CAP or smooth Voronoi-CAP. These observations are persistent in the other temporary anions we study below.

\section{Conclusion}
We have proposed in this work a more general criterion (referred to as the $\xi-$criterion, Eq. \eqref{eq:xi_criterion}) for choosing CAP parameters which minimize the CAP reflections. The $\xi$ metric gauges simultaneously if the VAE and width are artifacts of the CAP: the closer $\xi$ is to zero the better, for $\xi=0$ implies no CAP reflections (assuming all the CAP parameters are non zero). By plotting $\xi$ vs. the CAP parameter varied, one can easily locate the optimal CAP parameter in question by looking at the minima of $\xi$. It is worth bearing in mind that different minima points in a $\xi$ plot may correspond to different resonances. 

Even though $\xi$ can in principle take any positive real number, it's typical range is $0 \leq \xi \leq 1$.  The $\xi-$criterion may be used in optimizing both box-CAPs or smooth Voronoi CAPs. It may be used to optimize any CAP parameter of our choosing -- be it $\eta$ or any of the spatial parameters. This is an important advantage of the $\xi-$criterion, given that it allows us to optimize any CAP parameter, using the same criterion. This important flexibility offered by the $\xi-$criterion notwithstanding, we have also shown in this work that optimizing the spatial parameters of the CAP appears to be more important than optimizing $\eta$, the latter having been the primary focus of conventional optimizations of the CAP (together with the minimum log-velocity criterion, Eq. \eqref{eq:classic_criterion}), whereby one fixes the spatial parameters and varies $\eta$ to see what value of $\eta$ corresponds to a minimum of the CAP log-velocity. Indeed, if the spatial parameters of the CAP are not right, then an $\eta-$trajectory won't yield an acceptable minimization of the CAP reflections (or one may not even find any resonances, to begin with). In other words, the effectiveness of an $\eta-$trajectory is highly dependent on the fixed values of the spatial parameters. For this reason, we have proposed in this paper to fix $\eta$ and vary the spatial parameters of the CAP ($r^o_\alpha-$trajectories) to minimize the CAP reflections.

We have used the shape resonances of $^2\Pi_g$ \ce{N2-} state to illustrate how one can combine  $r^o_\alpha-$trajectories and the $\xi-$criterion in minimizing CAP reflections and determining resonance energies and widths of temporary anions. We studied the resonance by running spatial trajectories at four different fixed $\eta$ values -- namely, $\eta=0.00100$a.u., $0.00500$a.u., $0.01000$a.u., and $0.02000$a.u..  --  at different levels of theories (HF or EOM-EA-CCSD), and using different CAP functionals (box-CAP or smooth Voronoi -CAP). In general, we observed that fixing $\eta$ as low as $\eta=0.00100$a.u., we rarely found spatial parameters where the CAP reflections were sufficiently minimized. This was the case at all levels of theory and independent of the type of CAP. On the other hand, we were always able to locate at least one set of spatial parameters in the spatial trajectories at which the CAP reflections are sufficiently minimized (i.e., with $\xi \leq 0.05$) when we fixed $\eta$ at $\eta=0.00500$a.u., $0.01000$a.u., or $0.02000$a.u.. Thus, we recommend setting $\eta$ to at least $\eta=0.00500$a.u. when running $r^o_\alpha-$trajectories. Moreover, in the case of box-CAP $r^o_\alpha-$trajectories, it is best to choose a scheme to vary the spatial parameters which ensures that the point symmetry of resonant states is not broken in the presence of the CAP.

%The results presented in this paper also go on to show that it is possible to locate more than one resonance in a single $r^o_\alpha-$trajectory, using the $\xi-$criterion, Eq. \ref{eq:xi_criterion}.

We also observed bifurcations in the $r^o-$trajectories computed at the box- and smooth Voronoi CAP/HF levels for the temporary anions of \ce{N2-}. The asymptotic limits of these bifurcations are characterized by no reflections, i.e., they have $\Im\left<-i \eta W \right> = \Re\left<-i \eta W \right> = 0 $, and naturally, $\xi=0$.

It is worth noting from Table \ref{tbl:dinitrogen} that, for a fixed $\eta$, the optimal CAP box dimension according to the $\xi-$criterion, does not change that much between HF and EOM-EA-CCSD, and the same applies to the smooth Voronoi-CAP. 

In addition, optimal $r^o$ is approximately the same, for fixed $\eta$, whether we use box-CAP or smooth Voronoi-CAP. This would suggest an efficient strategy for optimizing the CAP spatial parameters if one chooses to use the spatial variation scheme used in this paper (i.e., for box-CAP, if one chooses to set $r^o_x=r^o_y=r^o_z$ and gradually varies the dimension of the cube) for it appears one could optimize the spatial parameters at a low level of theory like CAP-HF, followed by an optimization of the same spatial parameters (or even $\eta$) at a higher level (like CAP-EOM-EA-CCSD) in a small interval around the optimal parameters from the lower level of theory.

% If you have acknowledgments, this puts in the proper section head.
\begin{acknowledgments}
Funding from the European Research Council (ERC) under the European Union's Horizon 2020 
research and innovation program (Grant Agreement No. 851766) is gratefully acknowledged. 
Resources and services used in this work were partly provided by the VSC (Flemish Supercomputer 
Center), funded by the Research Foundation - Flanders (FWO) and the Flemish Government.
\end{acknowledgments}

% Create the reference section using BibTeX:
\bibliography{Resonances_CAP}

\end{document}

%\preprint{AIP/123-QED}

\title[Suppl. Mat.: CAP Analysis and Optimization]{Supplementary Material: Analysis and Optimization of Resonance Energies and Widths Using Complex Absorbing Potentials}
% Force line breaks with \\
\author{J.A Gyamfi}
 \email{jerrymanappiahene.gyamfi@kuleuven.be}
 \affiliation{Department of Chemistry, KU Leuven, Celestijnenlaan 200F, B-3001 Leuven, Belgium}%Lines break automatically or can be forced with \\
\author{T.-C. Jagau}%
 \email{thomas.jagau@kuleuven.be}
\affiliation{Department of Chemistry, KU Leuven, Celestijnenlaan 200F, B-3001 Leuven, Belgium}%

\date{\today}% It is always \today, today,
             %  but any date may be explicitly specified

\maketitle
\section{Short introduction to quadratic box-CAP and smooth Voronoi CAP}
For a quadratic box-CAP, the operator $W$ is defined as,
			
			\begin{subequations} \label{eq:W_alpha_quadratic}
			\begin{align}
			W & = \sum_{\alpha = x,y,z} \ W_\alpha \\
			W_\alpha & = \begin{cases} (\Delta_\alpha - r^o_\alpha)^2 \ & \text{if } \Delta_\alpha >r^o_\alpha \\ 
			0 \ & \text{otherwise }
			\end{cases} \label{eq:W_alpha_quadratic_b}\\
		   \Delta_\alpha & \equiv \abs{r_\alpha - o_\alpha}
			\end{align}
			\end{subequations}
where $r_\alpha$ is the electronic coordinate along axis $\alpha$. The vector $(o_x,o_y,o_z)$ is the origin of the CAP, which, for example, in the Q-Chem\cite{qchem50}
package, is chosen to coincide with the center of nuclear charge\cite{benda17}, i.e.,
			\begin{equation}
			o_\alpha = \frac{\sum_k R_{k,\alpha} Z_k}{\sum_k Z_k} \ ,
			\end{equation}
where $Z_k$ and $R_k$ are the charge and coordinate of the $k-$th nucleus, respectively. 

The parameters $\{r^o_\alpha\}$ in Eq. \eqref{eq:W_alpha_quadratic_b} determine a cuboid (or 'CAP box'), the outside of which defines the absorption region in space where we wish to artificially damp the electronic wavefunction; inside this cuboid, the CAP is not applied. Choosing appropriate CAP box size is important and also tricky: too small the box, the higher the CAP reflections, but if the box is too big, the CAP will fail to intercept the electronic wavefunction and the result will be as if the CAP was not even applied.

In the case of smooth Voronoi CAP, the boundary separating the absorbing region from the non-absorbing region is no longer a box but conforms more to the molecule's geometry. 
The operator $W(\va{r})$ is defined as
			\begin{equation}
			W (\va{r})= \begin{cases}
			(r_{av}(\va{r}) - r^o)^2, & \ \text{if } \bar{r} \leq r^o\\
			0, & \ \text{otherwise}\\
			\end{cases}
			\end{equation}
where $r_{av}(\va{r})$ is a weighted geometric average of electron-nuclei distances defined as
			\begin{equation}
			r_{av}(\va{r}) = \sqrt{\frac{\sum_I w_I \cdot r_I(\va{r})^2}{\sum_I w_I(\va{r})}}
			\end{equation}
where the index $I$ runs over all nuclei. The weight $w_I(\va{r})$ is defined as
			\begin{equation}
			w_I(\va{r}) \equiv \frac{1}{\left[r_I(\va{r})^2 - r_{nearest}(\va{r})^2 +\left( 1\text{a.u.}\right)^2\right]^2}
			\end{equation}
and
			\begin{equation}
			r_{nearest}(\va{r}) \equiv \min_I\left( \abs{\va{r} - \va{R}_I} \right)
			\end{equation}

\section{Geometries used for the $r^o-$trajectories (in \r{A}ngstroms).}

\subsection{\ce{N2-}:}
	\begin{verbatim}
    I     Atom           X                Y                Z
 ----------------------------------------------------------------
    1      N       0.0000000000     0.0000000000    -0.5488500000
    2      N       0.0000000000     0.0000000000     0.5488500000
 ----------------------------------------------------------------	
	\end{verbatim}

\bibliography{Resonances_CAP}